\documentstyle[12pt]{article}
\begin{document}

%DEFINITIONS

\newcommand\balpha{\mbox{\boldmath $\alpha$}}
\newcommand\bbeta{\mbox{\boldmath $\beta$}}
\newcommand\bgamma{\mbox{\boldmath $\gamma$}}
\newcommand\bomega{\mbox{\boldmath $\omega$}}
\newcommand\blambda{\mbox{\boldmath $\lambda$}}
\newcommand\bmu{\mbox{\boldmath $\mu$}}
\newcommand\bphi{\mbox{\boldmath $\phi$}}
\newcommand\bzeta{\mbox{\boldmath $\z$}}
\newcommand\bsigma{\mbox{\boldmath $\sigma$}}
\newcommand\bepsilon{\mbox{\boldmath $\epsilon$}}

\newcommand{\be}{\begin{eqnarray}}
\newcommand{\ee}{\end{eqnarray}}
\newcommand{\nn}{\nonumber}

\newcommand{\ft}[2]{{\textstyle\frac{#1}{#2}}}
\newcommand{\eqn}[1]{(\ref{#1})}
\newcommand{\vsone}{\vspace{1cm}}
 
\begin{titlepage}

\begin{flushright}
TIFR/TH/99-03 \\
hep-th/yymmddd
\end{flushright}
\begin{centering}
\vspace{.2in}
{\Large {\bf A Note on $1/4$-BPS States}}\\
\vspace{.4in}
David Tong${}$ \\
\vspace{.4in}
Tata Institute of Fundamental Research, \\
Homi Bhabha Rd, Mumbai, 400 005, India\\
{\tt tong@theory.tifr.res.in}\\
\vspace{.6in}
{\bf Abstract} \\
\end{centering}
\vspace{.1in}
We study classical solutions of ${\cal N}=4$ super Yang-Mills 
theories that are invariant under $1/4$ of the supersymmetry 
generators. Expressions for the mass and electric charge 
of the configurations are derived as functions on the monopole 
moduli space. These functions also provide a method of 
determining the number of normalisable bosonic zero modes.

%\vspace{.05in}

%\baselineskip=.3in
\end{titlepage}

\subsubsection*{Introduction}
The realisation of field theories as the low-energy 
world-volume dynamics of configurations of branes in type II 
string theory and M-theory has led to simple 
geometrical explanations of many field theoretic phenomena. 
The archetypal example is that 
of ${\cal N}=4$ super Yang-Mills which is the world-volume 
theory of D3-branes. 
On the Coulomb branch, $(p,q)$ 
strings stretched between each pair of D3-branes correctly 
reproduces the expected field theory spectrum of electric, 
magnetic and dyonic $1/2$-BPS states 
\cite{witten}. For three or more D3-branes however, 
there exists the possibility of suspending between them  
networks of $(p,q)$ strings constructed from three-string 
junctions \cite{disc}. If the D3-branes are planar then such 
networks preserve $1/4$ of the 
space-time supersymmetry \cite{sunil,sen,rey} and correspond 
to dyonic $1/4$-BPS states of the Yang-Mills theory 
\cite{berg,bergkol} thus  
leading to a much richer spectrum of BPS states than previously 
suspected for gauge groups of rank greater than one. The distinctive 
signature of such states is 
that the magnetic and electric charge vectors are not  
parallel \cite{ct,berg}.
These states persist at 
weak coupling in the field theory where they are amenable to 
standard semi-classical analysis and, in particular, one 
expects classical solutions with non-parallel magnetic 
and electric charges to exist.

The equations of motion governing such $1/4$-BPS configurations 
consist of the usual monopole Bogomol'nyi equation, together with 
a second order equation for an adjoint scalar  
field derived from Gauss' law. 
Explicit solutions to these equations  were first given 
in \cite{hhs,kawo} using a spherically symmetric 
ansatz. For the special case of the $(1,1)$ dyon of $SU(3)$ 
gauge group, solutions have also been found using Nahm data 
\cite{leeyi}. Moreover, as we shall review 
below, the authors of \cite{leeyi} also demonstrated that 
the space of gauge invariant $1/4$-BPS solutions with given 
magnetic charge is determined by the moduli space of $1/2$-BPS 
monopole solutions of the same charge. These monopole moduli 
spaces have been studied extensively in the 
past and many of their properties are understood \cite{ah}. 
However, although the existence of $1/4$-BPS solutions 
can be shown, little else is known about these configurations. 
In particular, neither the mass nor electric charge can be 
determined without an asymptotic solution to Gauss' law.

In this note, we derive an expression for both the 
mass and electric charge of $1/4$-BPS states as a 
function over the monopole moduli space. The former 
is shown to be equal to the norm squared of a 
particular Killing vector on the moduli space. 
We demonstrate this 
technique using the $(1,1)$ dyon of 
$SU(3)$ as an example and reproduce the results of 
\cite{leeyi}. 
Finally, we discuss the moduli space relevant to the low-energy 
dynamics of these $1/4$-BPS states. The existence of non-normalisable 
zero-modes means this space differs from the moduli space of 
solutions. We use the expression for the electric charge to 
determine the number of normalisable bosonic zero modes. As this work 
was completed, the preprint \cite{bhlms} appeared, where the number 
of normalisable bosonic zero modes is also calculated, togther with 
the normalisable fermionic zero modes. 

\subsubsection*{Bogomol'nyi Equations}

We start by reviewing the derivation of the Bogomol'nyi 
equations \cite{ct,kawo,leeyi} in ${\cal N}=4$ 
super Yang-Mills theories with arbitrary simple gauge group, 
${\cal G}$. 
It is simple to show \cite{leeyi} that preservation of $1/4$ 
supersymmetry requires four of the scalar fields to 
vanish, reflecting the fact that the corresponding BPS string 
network is planar. Our starting point 
is therefore the theory with only two real adjoint scalars  
and the configurations we discuss will also be 
solutions of the theory with ${\cal N}=2$ supersymmetry. The bosonic 
Hamiltonian reads,
\be
H=\ft12 {\rm Tr }\int{\rm d}^3x\ \left\{
(E_i)^2+(B_i)^2+({\cal D}_0\phi_1)^2 
+({\cal D}_0\phi_2)^2+({\cal D}_i\phi_1)^2
+({\cal D}_i\phi_2)^2-e^2[\phi_1,\phi_2]^2\right\}
\nn\ee
where $i=1,2,3$ is a spatial index and $e$ is the Yang-Mills 
coupling constant. As usual, the Bogomol'nyi 
equations are derived by completing the square. Defining,
\be
a&=&\cos\alpha\,\phi_1-\sin\alpha\,\phi_2 \nn\\
b&=&\sin\alpha\,\phi_1+\cos\alpha\,\phi_2
\label{alpha}\ee
for some angle $\alpha$, the Hamiltonian may be rewritten as,
\be
H&=& \ft12{\rm Tr}\int{\rm d}^3x\ \left\{
(E_i-{\cal D}_ia)^2+(B_i-{\cal D}_ib)^2
+({\cal D}_0a)^2+({\cal D}_0b)^2 \right. \nn\\
&& \qquad\qquad\qquad \left.-\ e^2[a,b]^2+2E_i{\cal D}_ia
+2B_i{\cal D}_ib\right\} \nn\\ 
&=&\ft12{\rm Tr }\int{\rm d}^3x\ \left\{(E_i-{\cal D}_ia)^2+
(B_i-{\cal D}_ib)^2+({\cal D}_0a)^2 \right.\label{bound}\\
&&\qquad\qquad\qquad \left. +\ ({\cal D}_0b-ie[a,b])^2
+2\partial_i\left(aE_i+bB_i\right)\right\} \nn\\
&\geq& {\bf a}\cdot{\bf q}+{\bf b}\cdot{\bf g}
\nn\ee
where ${\bf q}$ and ${\bf g}$ are the electric and magnetic 
charges respectively. Both are vectors of dimension 
$r={\rm rank}\,({\cal G})$. The vacuum 
expectation value (VEV) of $a$ is taken 
to lie in the Cartan subalgebra, 
$\langle a\rangle ={\bf a}\cdot{\bf H}$, with a similar 
expression for $b$. Converting the last two terms of the first line 
of \eqn{bound} to surface integrals requires use of the 
Bianchi identity, ${\cal D}_iB_i=0$, together with  
the equation of motion for $A_0$, usually referred to as 
Gauss' law,
\be
{\cal D}_iE_i=ie[a,{\cal D}_0a]+ie[b,{\cal D}_0b]\ .
\label{gauss}\ee
The inequality \eqn{bound} is saturated by time-independent solutions 
$(\partial_0=0)$ with the time-component of the gauge field 
given by $A_0=-a$ and the remaining gauge fields and adjoint 
scalar, $b$, determined by the usual Bogomol'nyi equation, 
\be
B_i&=&{\cal D}_ib\ .
\label{bog}\ee
Finally, Gauss' law requires the adjoint scalar, $a$, to obey,
\be
{\cal D}^2a-e^2[b,[b,a]]=0\ .
\label{d2a}\ee
Maximising the Hamiltonian bound \eqn{bound} with respect to 
the parameter $\alpha$ yields the equality 
${\bf a}\cdot{\bf g}={\bf b}\cdot{\bf q}$.
This equation is automatically satisfied by any solution to 
\eqn{bog} and \eqn{d2a} as can be seen by a simple 
integration by parts. 

When the electric charge vector, ${\bf q}$,  
is parallel to the magnetic charge vector, ${\bf g}$, the 
resulting configuration is a dyon solution which breaks 
$1/2$ of the supersymmetry generators. The spectrum of 
such states has been exhaustively investigated over the 
past few years. When the electric 
and magnetic charge vectors are not parallel however, the configuration 
breaks $3/4$ of the supersymmetry generators \cite{ct}. 
Such configurations correspond to three-string junctions 
\cite{berg,bergkol} and determining the spectrum of states in the 
field theory remains an open problem.

Solutions to \eqn{bog} have been studied in great detail. 
The VEV ${\bf b}$ selects $r$ simple roots\footnote{We 
assume ${\bf b}$ lies strictly within a Weyl chamber. 
This is stronger than 
the usual requirement that the gauge group is broken to its maximal 
torus}, $\bbeta_a$, $a=1,\ldots,r$. Topological considerations 
require the magnetic charge vector to be a linear combination 
of the co-roots: ${\bf g}=(4\pi /e)\sum_an^a\bbeta_a^\star$, 
for integers $n^a$. The moduli space, ${\cal M}_{\bf g}$, 
of gauge independent, purely magnetic solutions with 
magnetic charge, ${\bf g}$ and mass ${\bf b}\cdot{\bf g}$, 
has dimension $m=4\sum_an^a$ \cite{wein}. 
Let $X^p$, $p=1,\ldots,m$, denote coordinates on ${\cal M}_{\bf g}$. 
Tangent vectors are identified with zero modes, 
$(\delta_pA_i,\delta_pb)$, given by solutions to the linearised 
Bogomol'nyi equation  together with a suitable gauge fixing condition, 
taken to be background gauge,
\be 
&&\epsilon_{ijk}{\cal D}_j\delta_p A_k-{\cal D}_i\delta_p b
+ie[\delta_p A_i,b]=0 \nn\\
&&\hspace{.4in}{\cal D}_i\delta_p A_i-ie[b,\delta_p b]=0\ .
\label{gf}\ee
The overlap of zero modes defines a natural, complete, hyperK\"ahler 
metric on ${\cal M}_{\bf g}$ given by,
\be
g_{pq}={\rm Tr}\int{\rm d}^3x\ \left(\delta_pA_i\,\delta_qA_i+
\delta_pb\,\delta_qb\right)\ .
\label{metric}\ee
A subset of these zero modes are generated by bosonic 
symmetries of the vacuum which are broken by the monopole 
configuration, specifically translational, rotational and 
global abelian gauge symmetries. It is simple to see that each of 
these broken symmetries results in an isometry of the metric $g_{pq}$ 
and the corresponding zero modes are Killing vectors.  
Of particular interest will be the $h\leq{\rm min}(m/4,r)$ 
isometries generated by large gauge transformations of the 
form
\be
\delta_{\rm g.t.} A_i={\cal D}_i\epsilon\ \ \ \ ;\ \ \ \ 
\delta_{\rm g.t.} b=-ie[b,\epsilon] 
\label{gt}\ee
where the gauge parameter $\epsilon$ does not tend asymptotically 
to zero but to some element, $\bepsilon\cdot{\bf H}$, of the 
Cartan subalgebra. Such deformations automatically solve the linearised 
Bogomol'nyi equation but are zero modes only if they simultaneously 
satisfy the background gauge fixing condition \eqn{gf}. This  
requirement is recognised as equivalent to the equation 
of motion for the adjoint scalar $a$ \eqn{d2a}. 
The existence of a unique zero mode for each element of the Cartan 
subalgebra broken by the monopole configuration ensures a unique 
solution to \eqn{d2a} for each VEV ${\bf a}$ \cite{leeyi}. Therefore, 
${\cal M}_{\bf g}$ is also the moduli space of $1/4$-BPS dyon 
configurations.

\subsubsection*{A Potential on the Monopole Moduli Space}

In summary, for fixed value of $\alpha$, and fixed magnetic 
charge ${\bf g}$, there exists a one-to-one 
correspondence between solutions to \eqn{bog} describing 
purely magnetic configurations and solutions 
to \eqn{bog} and \eqn{d2a} describing $1/4$-BPS dyons.   
However, the electric charge, and therefore the mass, of the 
$1/4$-BPS dyon configurations remains undetermined. In general, these 
quantities are given by a function over ${\cal M}_{\bf g}$. 
One method to determine this function is to explicitly 
solve equation \eqn{d2a}, at least to lowest order 
in the radial spatial coordinate, $r$. The two equations 
$B_i={\cal D}_ib$ and $E_i={\cal D}_ia$ ensure the 
asymptotic behaviour of $b$ and $a$ encodes the 
magnetic and electric charges respectively, 
\be
b&=&{\bf b}\cdot{\bf H}-{\bf g}\cdot{\bf H}/4\pi r +O(1/r^2)\nn\\
a&=&{\bf a}\cdot{\bf H}-{\bf q}\cdot{\bf H}/4\pi r +O(1/r^2)
\label{asymp}\ee
For general monopole configurations, determining even the 
asymptotic form of $a$ is difficult. However, there 
exists a simple formal expression for both quantities in terms 
of the monopole moduli space metric \eqn{metric}. We start 
with the electric contribution to the mass, 
\be
{\bf a}\cdot{\bf q}&=&{\rm Tr}\int{\rm d}^3x\ \partial_i
\left(aE_i\right) ={\rm Tr}\int{\rm d}^3x\ \partial_i
\left(a{\cal D}_ia\right) \nn\\
&=&{\rm Tr}\int{\rm d}^3x\ 
\left(({\cal D}_ia)^2-e^2[a,b]^2\right)
\label{aq}\ee
where we have used the equation of motion \eqn{d2a}. As in the 
previous section, we can make progress by considering the 
subset of zero modes of the Bogomol'nyi equation that  
arise from large gauge transformations and the corresponding 
Killing vectors on ${\cal M}_{\bf g}$.  
Denote as $\bepsilon\cdot{\bf K}^p$ the components of the 
Killing vector on 
${\cal M}_{\bf g}$ that is generated by the gauge transformation 
\eqn{gt} that asymptotes to $\bepsilon\cdot{\bf H}$. Consider 
then the transformation with gauge parameter 
$\epsilon =a$, which is a zero mode courtesy 
of \eqn{d2a}. Expanding in a basis of tangent vectors, we have,
\be
\delta_{\rm g.t.}A_i={\cal D}_ia=({\bf a}\cdot{\bf K}^p)
\delta_pA_i \nn\\
\delta_{\rm g.t.}b=-ie[b,a]=({\bf a}\cdot{\bf K}^p)
\delta_pb
\nn\ee
which, after substitution into \eqn{aq} above, gives the promised 
expression for the electric contribution to the mass of the 
$1/4$-BPS dyon as the norm squared of a Killing vector 
on ${\cal M}_{\bf g}$, 
\be
{\bf a}\cdot{\bf q}=g_{pq}({\bf a}\cdot{\bf K}^p)({\bf a}\cdot{\bf K}^q)
\label{pot}\ee
This potential was first derived in the context of three-dimensional 
constrained instantons in \cite{cd}. The Killing vector 
${\bf a}\cdot{\bf K}^p$ is tri-holomorphic and curiously a 
sigma model with such a potential is of the form required 
to realise up to eight supercharges \cite{agf}. 
However, as is discussed in the final section, this model 
cannot be relevant to the low-energy dynamics 
of $1/4$-BPS states as it would result in a time 
dependent electric charge, ${\bf q}$. 

Finally, the expression for the electric charge may be derived in 
a similar manner. Repeating the manoeuvres that led to \eqn{aq}, we 
have,
\be
\bepsilon\cdot{\bf q}={\rm Tr}\int{\rm d}^3x\ 
\left({\cal D}_i\epsilon {\cal D}_ia-e^2[\epsilon ,b][a,b]\right)
\label{eq}\ee
for arbitrary vector $\bepsilon$. Once again expanding in 
tangent vectors, we find
\be
{\bf q}=g_{pq}({\bf a}\cdot{\bf K}^p)\ {\bf K}^q\ .
\nn\ee

\subsubsection*{An Example: The (1,1) SU(3) Dyon}

A simple, but trivial, use of the potential \eqn{pot} is in describing 
$1/2$-BPS dyons, even for $SU(2)$ gauge group. In this 
case, the potential is constant over the moduli space and 
can be easily shown to be equal to the mass difference 
between the monopole and dyon. 

The simplest non-trivial application of the potential is in  
the description of $1/4$-BPS configurations in an $SU(3)$ gauge 
group. The VEV ${\bf b}$ selects two simple roots, labeled 
$\balpha$ and $\bbeta$. Monopole configurations with magnetic charge 
${\bf g}=(4\pi /e)\balpha^\star$ or ${\bf g}=(4\pi /e)\bbeta^\star$ 
are known as 
``fundamental''. Both sectors have $4$ zero modes and monopole 
mass given by $m_1=(4\pi /e){\bf b}\cdot\balpha^\star$ and 
$m_2=(4\pi /e){\bf b}\cdot\bbeta^\star$. 
The potential on the monopole moduli space  
is always a constant and the resulting electric charge of any dyon 
is necessarily proportional to the magnetic charge. 

In order to construct 
a $1/4$-BPS configuration, we must consider the magnetic charge 
charge sector  ${\bf g}=(4\pi /e)(\balpha^\star+\bbeta^\star$). Purely 
magnetic solutions to \eqn{bog} have mass $m_1+m_2$ and an 
eight-dimensional moduli space generically of the form 
\cite{gl,lwy}
\be
{\cal M}_{(1,1)}=R^3\times\frac{R\times{\cal M}_{TN}}
{Z}
\nn\ee
The $R^3\times R$ factor is parametrised by collective coordinates 
${\vec x}$, corresponding to spatial translations, and $\chi$ 
corresponding to a global gauge transformation. The 
corresponding metric is flat,
\be
{\rm d}s^2_{flat}=(m_1+m_2){\rm d}{\vec x}^2+\frac{1}{m_1+m_2}
d\chi^2\ .
\nn\ee
The factor $M_{TN}$ is Taub-Nut space and can be thought of 
as describing the relative separation, ${\vec r}$, and 
relative phase $\psi$ of a pair of distinct fundamental 
monopoles. The metric on this space is given by,
\be
{\rm d}s^2_{TN}=\left(1+\frac{2(m_1+m_2)}{m_1m_2 r}\right)
{\rm d}{\vec r}+\frac{(m_1+m_2)^2}{4m_1^2m_2^2}
\left(1+\frac{2(m_1+m_2)}{m_1m_2 r}\right)^{-1}({\rm d}\psi
+\cos\theta\,{\rm d}\phi)^2
\nn\ee
where $\psi$ has period $4\pi$, ensuring the metric has no singularity 
at the origin. The discrete identification $Z$ acts as 
$(\chi,\psi)=(\chi+2\pi,\psi+4\pi m_2/(m_1+m_2))$. 
If the ratio $m_1/m_2$ is rational, $\chi$ becomes periodic and 
the $R$ factor collapses to $S^1$.  

A natural basis of gauge transformations in the Cartan 
subalgebra is given by
\be
{\bf Q}_1\cdot{\bf H}=\frac{2}{3}\left(2\balpha+\bbeta\right)
\cdot{\bf H}\ \ \ \ \ ;\ \ \ \ \ {\bf Q}_2\cdot{\bf H}
=\frac{2}{3}\left(\balpha+2\bbeta\right)\cdot{\bf H}
\nn\ee
both of which are integer valued and between them 
generate the two Killing vector fields 
$\partial /\partial_\chi$ and $\partial /\partial_\psi$ on 
${\cal M}_{(1,1)}$. The specific linear combination of charges 
which generate each isometry can be determined by an analysis 
of the low-energy dynamics of $1/2$-BPS configurations  
and is given by \cite{gl,lwy}, 
\be
{\bf Q}_\chi =\frac{m_1{\bf Q}_1+m_2{\bf Q}_2}{m_1+m_2}
\ \ \ \ \ ;\ \ \ \ \ {\bf Q}_\psi=\frac{1}{2}\left(
{\bf Q}_1-{\bf Q}_2\right)
\nn\ee
Notice that these two isometries of ${\cal M}_{\bf g}$ 
are not generated by orthogonal gauge transformations, a 
fact related to the non-compactness of the ${\bf R}$ 
factor of ${\cal M}_{(1,1)}$. We are now in a position 
to determine the electric charge of the $1/4$-BPS dyon 
solutions. Decomposing ${\bf a}$ in terms of the two 
generators above gives
\be
{\bf a}=\frac{e}{2\pi}\left({\bf a}\cdot{\bf g}\right){\bf Q}_\chi 
+\left({\bf a}\cdot(\balpha-\bbeta)-\frac{m_1-m_2}{m_1+m_2}
\frac{e}{2\pi}({\bf a}\cdot{\bf g})\right){\bf Q}_\psi
\nn\ee
which, when substituted in \eqn{pot}, gives the electric 
contribution to the mass of the $1/4$-BPS dyon configuration 
as
\be 
{\bf a}\cdot{\bf q}&=&g_{\chi\chi}\frac{e^2}{4\pi}
({\bf a}\cdot{\bf g})^2+g_{\psi\psi}
\left({\bf a}\cdot(\balpha-\bbeta)-\frac{m_1-m_2}{m_1+m_2}
\frac{e}{2\pi}({\bf a}\cdot{\bf g})\right)^2 \nn\\ 
&& \qquad +
g_{\chi\psi}\frac{e}{\pi}({\bf a}\cdot{\bf g})\left({\bf a}
\cdot(\balpha-\bbeta)
-\frac{m_1-m_2}{m_1+m_2}\frac{e}{2\pi}({\bf a}\cdot{\bf g})\right)\nn\\
&=&\frac{e}{2\pi}\frac{({\bf a}\cdot{\bf g})^2}{({\bf b}\cdot{\bf g})\ }
+\frac{(m_1+m_2)r}{m_1+m_2+2m_1m_2r}\left({\bf a}\cdot
(\balpha-\bbeta)-\frac{m_1-m_2}{m_1+m_2}
\frac{e}{2\pi}({\bf a}\cdot{\bf g})\right)^2
\nn\ee
The electric charge of the configuration is now simply determined 
given the two non-parallel components 
${\bf a}\cdot{\bf q}$ and ${\bf b}\cdot{\bf q}={\bf a}\cdot{\bf g}$. 
After a little algebra we find, 
\be
{\bf q}= \left(\frac{{\bf a}\cdot{\bf g}}{{\bf b}\cdot{\bf g}}\right)
(\balpha +\bbeta ) + \Delta q\left({\bf a}-
\frac{({\bf a}\cdot{\bf b})}{|{\bf b}|^2}{\bf b}\right)
\nn\ee
The first term is common to all dyons, including 
those preserving $1/2$ supersymmetry. The second term is non-zero 
only for $1/4$-BPS dyons where $\Delta q$, an order parameter 
for the breaking of the extra supersymmetry generators, is 
explicitly given by, 
\be
\Delta q&=&
\frac{|{\bf b}|}{|{\bf a}||{\bf b}|-({\bf a}\cdot{\bf b})}
\frac{(m_1+m_2)r}{m_1+m_2+2m_1m_2r}\left({\bf a}\cdot({\balpha-\bbeta})
-\frac{m_1-m_2}{m_1+m_2}\frac{e}{2\pi}({\bf a}\cdot{\bf g})\right)^2
\nn\\
&=&\frac{4(m_1^2+m_2^2+m_1m_2)r}{(m_1+m_2)(m_1+m_2+2m_1m_2r)}
\label{delq}\ee
This expression for the electric charge of a $(1,1)$ dyon as a 
function on the monopole moduli space was also obtained by Lee and Yi 
\cite{leeyi} (see equation (3.11) of this reference) using a different 
technique.

Although we have determined the potential only in the simple 
case of the $(1,1)$ dyon, several features will persist for 
general magnetic charges. 
From \eqn{delq}, we see that at the point $r=0\,$ of 
${\cal M}_{(1,1)}$ the electric charge is proportional to the 
magnetic charge and the dyon preserves $1/2$ of the supersymmetries. 
A similar situation occurs for any configuration with 
${\bf g}=(4\pi /e)k\balpha^\star$ for some root $\balpha$ and 
integer $k$. The relevant 
monopole moduli spaces contain the 
$SU(2)$ monopole moduli spaces of charge $k$ as a 
submanifold of fixed points of the relative $U(1)$ isometries. 
The corresponding configurations are simply the usual 
$1/2$-BPS dyons that exist for complex Higgs fields 
\cite{tim}. For 
${\bf g}$ not proportional to a root, the isometries 
have no fixed points and there are no $1/2$-BPS solutions. 
As $r\rightarrow\infty$, $\Delta q$ approaches a limiting value.  
The asymptotic forms of all monopole moduli spaces are known 
\cite{lwy2} and such behaviour is generic and is related 
to the instablity of the string network \cite{leeyi}.

\subsubsection*{Normalisable Zero Modes}

In this final section, we comment on the dynamics 
of the classical $1/4$-BPS configurations which, at low-energies, 
should be describable in a moduli space approximation. 
Locally, the moduli space of $1/4$-BPS dyon solutions of 
magnetic charge ${\bf g}$ is given by $S^1\times
{\cal M}_{\bf g}$, where the $S^1$ factor refers to the 
$\alpha$ parameter introduced in \eqn{alpha}. The VEV 
${\bf b}$, and therefore the metric on ${\cal M}_{\bf g}$, 
varies with $\alpha$. Moreover, at specific points ${\bf b}$ 
may cross the wall of a Weyl chamber and the dimension of 
${\cal M}_{\bf g}$ will change discontinously. 

Naively, one expects the low-energy motion to be restricted to 
surfaces of constant electric charge on ${\cal M}_{\bf g}$. To see 
how this arises, consider the natural metric describing $1/4$-BPS 
dyon dynamics,
\be
\tilde{g}_{pq}={\rm Tr}\int{\rm d}^3x\ \left(
\delta_pA_i\, \delta_qA_i+\delta_pb\,\delta_qp
+\delta_pa\,\delta_qa\right)
\nn\ee
which differs from the usual monopole metric \eqn{metric} by 
the addition of the overlap of zero modes of $a$ satisfying,
\be
{\cal D}^2\delta_pa-e^2[b,[b,\delta_pa]]=2ie\left([\delta_pA_i,
{\cal D}_ia]-ie[\delta_pb,[b,a]]\right)
\nn\ee
From the asymptotic form of the scalar fields \eqn{asymp}, 
we see that any 
zero mode of $a$ which induces a change in the 
either the VEV, ${\bf a}$, or the electric charge, 
${\bf q}$, is non-normalisable and the component of the 
metric $\tilde{g}_{pq}$ in this direction diverges. 
As expected therefore, motion on the moduli space of solutions 
is restricted to surfaces of constant $\alpha$ and ${\bf q}$ 
and the number of normalisable bosonic zero modes is equal  
to the dimension of surfaces of constant electric charge. 
This is simple to determine given the expressions for 
${\bf q}$ derived previously. For each of the $h$ abelian 
gauge symmetries broken by the dyon configuration, there exists 
a function $\bepsilon\cdot{\bf q}$ on the monopole moduli 
space \eqn{eq}. 
With the exception of ${\bf b}\cdot{\bf q}={\bf a}\cdot{\bf g}$, 
all of these functions vary independently over 
${\cal M}_{\bf g}$. The surface of constant electric charge is  
therefore given by the intersection of $h-1$ level surfaces and 
generically describes a submanifold 
of ${\cal M}_{\bf g}$ of dimension $m-h+1$. This result was also 
derived in \cite{bhlms} using similar methods. These 
authors further determine the number of fermionic zero modes to 
be given by $2m-4h+4$. 

In the example of the $(1,1)$ dyon considered 
in the previous section, $m=8$ and $h=2$ and the number of bosonic and 
fermionic zero modes is 7 and 12 respectively. In this case the 
latter are all generated by broken supersymmetry transformations while 
the former consist of all collective coordinates on 
${\cal M}_{(1,1)}$ with the exception of the relative separation, 
$r$. The appearance of an odd number of bosonic zero modes however 
should give us pause. On general grounds, one would expect the 
low-energy effective action describing dyon motion to reflect 
the four supercharges preserved by the classical configurations. 
Such a sigma model requires a K\"ahler target space. The number 
of normalisable zero modes derived above therefore appears to be 
in contradiction with the 
dictates of supersymmetry. Further, without a supersymmetric 
effective action, it is unclear how semi-classical quantisation 
could produce the correct multiplet struture of $1/4$-BPS states. 
Clearly a resolution of these issues is required before further 
progress in determining the full BPS spectrum of ${\cal N}=4$ 
gauge theories can be made.

\subsubsection*{Acknowledgements}

It is a pleasure to thank Keshav Dasgupta, Jerome Gauntlett, Tim 
Hollowood, Sudipta Mukherji, Sunil Mukhi and Naoki 
Sasakura for useful 
discussions and comments. I am also grateful to the 
Mehta Research Institute for hospitality while part of 
this work was completed.

\end{document}